\baselineskip=24pt
\magnification=1200
\noindent
\centerline {NEARLY 
CLOSED LOOPS IN BIOLOGICAL SYSTEMS AS ELECTROMAGNETIC RECEPTORS}

\centerline {D. Eichler}

\centerline {Dept. of Physics, Ben-Gurion University, Beer Sheva, ISRAEL}

\centerline {and}

\centerline {Institute for Theoretical Physics, University of California, Santa
Barbara, CA, 93106-4030}

\vskip .5in

 It is noted here that when a nearly closed loop in a biological
system, such as a self-synapsing (autapsing) 
neuron or mutually synapsing pair, is 
exposed to an AC magnetic field, the induced electric fields
in the insulating gaps can be many orders of
magnitude larger than the average values typically discussed in the
literature.$^{1,2}$  It is suggested that animal nervous systems might 
possibly be affected in selected spots 
by  man-made 
alternating magnetic fields at weaker levels than previously supposed.
 Radio and microwave radiation should 
be considered particularly suspect.
\vfill
\eject
%%%%%%%%%%%%%

The possible influence of "everyday" electromagnetic (EM)
fields  on living 
systems has been a topic of great interest in recent 
years, with 
many controversial claims and counterclaims in the
 epidemiological, 
laboratory and theoretical arenas. Of particular 
interest  has been the theoretical question of 
whether, at typical  levels experienced from household appliances, 
trains, etc. such fields 
could conceivably affect living systems at levels
 that 
 thermal electrical noise and molecular agitation
 might appear to override 
$^{1,2}$. The claim in  these 
references, 
which otherwise assume the 
conservative premise that EM 
fields are suspect until proven safe, is 
that fields that are weaker than those caused by thermal noise 
must be beyond suspicion. 
It is noted in this letter, however, that if the effective brightness
temperature of the alternating electromagnetic field  greatly 
exceeds the  temperature of the system  -i.e. if the 
thermally averaged electromagnetic 
energy    per cavity mode 
is much higher than  kT - it cannot be rigorously demonstrated that 
such fields are entirely safe.

 Low frequency electric fields do not, on average, penetrate 
biological tissue 
effectively $^{1,2,3}$. 
Magnetic fields B, which penetrate effectively, exert a weaker 
instantaneous  force on 
individual charged particles as  compared to electric 
fields E of the same magnitude (or, equivalently, the same energy density)
by a factor of order (v/c), which is much less than unity. 
%, $(ev/c) X B$, where e is the charge, 
Here v is the velocity of the charged particle and c is the speed of light. 
%is small compared to the electric force eE.
The time averaged effect 
could of course be a 
more complicated matter, but this letter in any case focuses on electric fields.
Induced electric fields from alternating magnetic fields (AMF) are
limited in their curl by Faraday's law but the electric potential
gradient induced by the AMF, via the rearrangement of charges, is not.

Consider a nearly complete, circular loop of electrically conducting
material of radius $R$. Assume that the loop contains an insulating notch
(a slot gap) of length $\delta$. When the loop is
exposed to a time varying magnetic
field $Be^{i\omega t}$, the electromotive force around the loop is
$$
\oint E\cdot d\ell = {i\omega\over c}\pi R^{2} Be^{i\omega t}
\eqno(1)
$$
(We have neglected spatial dependence of the field assuming it to be
small on the scale $R$. We have also assumed that the magnetic field
 penetrates the biological tissue 
at  frequencies well below 1000 Mhz,
 as is generally
recognized to be the case, and for higher frequencies we  
note that the arguments presented here may apply only within a penetration 
depth of the surface of the organism. Unless otherwise stated,
we use cgs units in which $E$ and $B$ have the
same dimensionality, and remind the reader that in 
  free space 
 a 1 Gauss magnetic 
field has the same energy density as a 1 statvolt/cm electric field, 
that the statvolt/cm and Gauss have the same physical dimensionality 
in the primary dimensions of mass, space and time, and that 
1 statvolt/cm 
 = 300 V/cm.

If the rate of charge redistribution is larger than $\omega$, the
contribution to the integral is dominated by the gap, so that the peak
electric field in the gap is
$$
E = {\omega\over c}\, {\pi R^{2}\over\delta} B.
\eqno(2)
$$

Now consider a nearly closed biological loop of connected conducting
solution, e.g., a neuron of length $2\pi
R$ that closes upon itself, making a
synapse of width $\delta$ between its axon and one of its own
dendrites. A     geometric idealization is illustrated  in the 
figure. The resistance across the synapse is probably dominated by the
neuronal membrane on either side of it (i.e. where the dashed and dotted line 
intersect),
which
has a surface resistivity of order $10^4$ ohms cm$^2$. Because of 
ionic currents and  polarization 
 in the extracellular fluid at the gap,  any 
potential difference across the synapse is likely to occur mostly in the 
membrane itself (where in fact its biological significance is 
possibly the greatest), and with this understanding we shall 
for brevity refer to it as the potential difference across the gap. 
The effective value of $\delta$ to be used in equation (2) may in fact be 
only twice the membrane thickness, and this reduction in $\delta$ 
will only serve to strengthen the arguments presented here.  Assuming a length
scale of order several hundred Angstroms, and the area of the synapse to
be the square of that, the resistance of the synapse is then of order
$10^{15}$ ohms, and easily dominates the total resistance of the
circuit. The EMF generated around the loop is then mostly in the synapse.
An
electric field of
$$
E=\left({\omega\over 10^{4}{\rm hz}}\right)\left({B\over 1{\rm
Gauss}}\right)\left({R \over 1 {\rm cm}}\right)^2\left({10^{-6}{\rm
cm}\over\delta}\right)\times 314{V\over{\rm cm}}
\eqno(3)
$$
is created at the synapse. Independent of the width of the synapse, a
voltage of
$$
V=0.3{\rm mV} \left({\omega\over 10^{4}{\rm hz}}\right)\left({B\over
1{\rm Gauss}}\right)\left({R\over 1{\rm cm}}\right)^2
\eqno(4)
$$
is created across it. 
  This is similar to the result obtained by Polk (1992) except that
he assumes a loop composed
 of small individual cells so 
that there are approximately 3000 individual gap
 junctions in a loop of 1 cm radius.  Thus, the electric fields
envisioned across a self-synapse can be larger by 
a factor of 3000 or so than in Polk (1992).  
While $R$ is likely to be less than 1 cm for a
given neuron, it is reasonable
that embedded in  neural nets are large loops containing  N synapses in
series (which I term an 
 N-cyclic synapse sequence, 
N at least two), and that $({R\over 1{\rm cm}})\gg N$. In
this case $E \gg \left({\omega\over 10^{4}{\rm hz}}\right)\left({B\over
1{\rm Gauss}}\right)N^{-1}$ statvolt/cm.

The maximum $\omega$ for which the above is valid is the inverse time
scale $\left({1\over ZC}\right)$ for charge redistribution around the
loop where $Z$ is the electrical resistance of the intraneuronal 
fluid (excluding the membrane) and $C$ is the capacitance of
the gap 
(including the membrane). 
If the loop has a constant cross-section, then $\left({1\over
ZC}\right)$ is ${\delta\over
2\pi R} \omega_{p}^{2}\tau_c$, where $\omega_p^2 =  4\pi nq^2/m$ 
and $\tau_c$ is the collision time. Here the quantities n, q and m are the 
number density, charge and mass of the dominant charge carriers. By equation
(3), the maximum electric field is
thus
$$
\eqalign {E_{max}&={\left({\delta\over 2\pi
R}\omega_{p}^{2}\tau_{c}\right)\over c\delta} \pi R^{2}B{\eta}\cr
&={\omega_{p}^{2}\tau_{c}R\over 2c}B{\eta}}
\eqno(5)
$$
The above assumes that the conducting medium of which the loop is
composed is collisional, {\it i.e.}, $\omega\tau_{c}<1$, which is usually
the case. For a dissipationless loop the right hand side of equation (5)
would be replaced by ${\omega_{p}\over c}RB$. A typical value of the
conductivity $\omega_{p}^{2}\tau_{c}$ in
biological fluid is $3\times 10^{10}s^{-1}$, or equivalently, 1.4 mho/m =
1.4 S/m. For intraneuronal fluid in a squid,  it can exceed 3 S/m, whereas for 
mammals, a more typical value is 0.8 S/m.$^5$  
In equation (5), $\eta$ is the effective ratio of cross-sectional areas
between the loop -- over most of its length -- and the gap. One expects that
  $\eta$ is large. This allows a faster
charge redistribution by a factor $\eta$. Estimates for $\eta$ in various
physiological networks are beyond the scope of this letter. Large neurons
 can have cross-sectional radii as high as 10 microns,
whereas gap scales could be as small as $10^2$ angstroms, so $\eta$ could
get as high as $10^6$. For large arteries, which  have cross sectional 
areas as large as millimeters, $\eta$ could conceivably be as high as 
$10^{10}$. Naively, this
extends the validity of equations (3) and (4) to include radio and possibly
microwave and IR frequencies. However, the overall resistance of the loop
depends the hierarchical branching pattern of the network it is a part
of, and needs to be evaluated on a case by case basis.

%A source of $P_{\rm kw}$ kilowatts of low frequency put into a solid
%angle of $\Omega$ creates a magnetic field at a distance of $D_{\rm cm}$
%cm of
%$$
%B=0.6\,D_{\rm cm}^{-1}P_{\rm kw}^{-1/2}
%\left({4\pi\over\Omega}\right)^{1/2}{\rm Gauss}
%\eqno(5)
%$$
The radii of loops, and the effective slot gap widths $\delta$ within
such loops that can exist in large mammals (e.g. humans) is a non-trivial
physiological question that is well beyond the scope of this letter.
Loops many (of order 10, say) cm in radius may exist in the circulatory and
nervous systems. The possibility of multiple coiling, as in a sparkplug,
would further extend
the range in 
the effective value of the loop area. By equation (3) and subsequent discussion,
electric fields could be
created in slot gaps that are even greater in magnitude than
 the alternating magnetic fields if the latter are applied at
frequencies
near  $\omega\sim{\omega_{p}^{2}\tau_{c}\delta\over
2\pi R}\sim 5\times 10^{3}\left({\delta\over 10^{-6}{\rm
cm}}\right)\left({1{\rm cm}\over R}\right){\rm hz}$.

If the loop is a bacterial chromosome, or any large, closed
 polymeric molecule, we suggest in a highly speculative spirit that 
there exists the
possibility of persistent (dissipationless) current being excited.
Defects in the polymer could act as slot gaps. In this case, the 
frequency of collective electronic oscillations could 
be of order $10^{15-16}$s$^{-1}$. 
 If the scale of the polymer is $10^{-5}$ cm,
and the defect of order 1\AA,  electric fields at the defect
could be comparable to
applied alternating magnetic fields at infrared  frequencies, and in
any case higher than the average value allowed by Faraday's law. Though 
infrared light would not penetrate deeply into biological tissue, it may 
nevertheless be an issue for, say, epidermal cells.

Field strengths of order of several Gauss, and frequencies $\omega$
 of order
$10^3$, which obtain       
in some  household appliances,
 can
give EMF's of order a  millivolt  over a scale of several cm.  Sources of 
high frequency RF, such as magnetic 
resonance imaging (MRI) devices, celluar phones, 
and microwave ovens are especially suspect. 
 Consider microwave emission at about 
2500 Mhz at the standard safety ceiling of 1 milliwatt cm$^{-2}$.
This corresponds to a magnetic field strength in air (essentially 
free space) of about 2 milligauss. The penetration depth into 
biological tissue of waves at this frequency is of 
order 1 cm. 
The EMF generated in a loop of 1 cm radius is about 4$\times 10^{-3}$ gauss cm, 
or about 1 volt.  If this EMF is concentrated across a single or even several 
neuronal membranes, its physical significance  could not be safely dismissed, 
as it is more than an order of magnitude more than the natural range 
of membrane potentials. Similar concerns would exist for the RF fields from  
magnetic resonance imaging (MRI) devices, and 
cellular phones at a distance of several cm,  where the  field can change at
a rate as high as $10^6$ gauss/s. 

It is not claimed here that even the maximal electric fields derived here
are dangerous, or that various claims of carcinogenic effects can be
accounted for via the electric field concentration mechanism discussed
here. We also note that the estimates$^{1,2}$ on the maximum average field
strength do not contradict the arguments presented here, and 
it is emphasized  that the field  attains significantly higher than average   
values only in small selected spots that comprise a small 
fraction of the total volume.
It is worth noting, however, that neural network viability in theoretical
models depends on sensitive balance between inhibition and excitation.
Slight systematic changes in the firing rates of large numbers of neurons
could qualitatively change the behavior of the net in a similar way
that, say, bubble chambers and cloud chambers are affected by very weak
perturbations. In each case, arguably, the system records information by
existing in a delicate state. It is also conceivable that self-synapsing 
neurons, because they can be efficient feedback loops, play a significant 
role in the "personality" of the neural net in which they are 
embedded.
Systematic interference in the function of self-synapsing neurons  
might therefore affect the global behavior of the network in addition 
to the chemistry of individual synapses.

In conclusion, the thermal noise limit needs to be used with caution when
applied as a safety guarantee for exposure to time varying magnetic
fields. The physiology of living systems may employ the same principles as
the meters and oscilliscopes that measure ``safely weak" fields while at
room temperature. Strictly speaking, the thermal noise threshold is not
necessarily established by the Johnson noise or thermal agitation $^{1,2,3}$ 
when the brightness temperature of the EM radiation
exceeds kT.

\noindent {\it note added:}
Autapsing neurons have been grown in culture and, though the autapses 
have   long been suspected to 
be an artifact of the culture growth,  it has very recently been reported
$^6$  that 
they are common in developing neocorticies of young rats. Most  neurons 
of the investigated class (level 5 pyramidal) 
 are found to autapse and 
have  on the average more than  two 
autapses per neuron. The spatial 
dimensions of the autapse circuit  can be as large as several hundred microns 
across.%, and the timescale for  an electrical impulses to propagate around 
%the loop are estimated to be $10^{-4}$s. 
 A  reasonable scaling to large mammals 
would be  that the spatial dimensions are roughly
 a factor 
of 3 to 20 larger.  This estimate is based on the fact that cell bodies 
 are  a factor of 3 or 4 larger in large mammals than in small rats, and 
the axonal lengths could scale more in proportion to the linear 
 size of the animal 
(E. White, private communication). It is conjectured
 here that cyclic sequences of synapses will eventually be 
found on still larger scales. 

I thank R.~Adair, J.~Arons, R. Belmaker, 
R.~Blandford, G. Erez, A.~Weisel-Eichler, P. 
Finklestein, F. Liebersat, A. Mizrahi,  M. Gutnick,  P. Israelovich, E.N.
Parker, S. Stokar and E.~White for
valuable conversations.  I thank the Institute of Theoretical Physics at
the University of California, Santa Barbara for its hospitality
while this paper was written. This work was supported in part by the 
U.S.
 Israel Binational Science Foundation.

\vskip .5in

\noindent
{\bf References}

\noindent
1. Adair, Robert K.
%``Constraints on biological effects of
%weak-extremely-low frequency electromagnetic fields."
{\it Phys.~Rev.~A}
{\bf 43}, 2 (1991).

\noindent
2. Bennet, William R. Jr. ``Cancer and Power Lines" in {\it Physics
Today}, April 1994, pp~23-9.

\noindent
3. Weaver, J.C. and Astumian, R.D. {\it Bioelectromagnetics Suppl}. {\bf
247}, 459 (1990)

\noindent 4. Polk, C.  {\it Bioelectromagnetics Suppl.} 1, 209 (1992)

\noindent 5. Kuffler, S. W., Nichols, J. G.,
 and Martin, A. R. "From Neuron to Brain"   (Sinauer Assoc.
 Inc., Sunderland, MA) 2nd edition (1984)

\noindent 6. Lubke, J. Markram, H.  Frotscher, M. 
and Sakmann, B.  {\it Journal of Neuroscience} 16, 3209 (1996)
\vfill
\eject
Firgure Caption: The idealized representation used in the text 
of a self-synapsing neuron or 
other nearly closed looop is illustrated. The dotted line denotes the 
circuit of integration in equation (1). The solid line denotes the 
neuronal membrane. The figure 
is not drawn to scale and the solid boundary should
in fact comprise a significant fraction of the gap.
\end